\documentclass[aps,twocolumn,nofootinbib]{revtex4-1}

\usepackage{amsmath,amssymb,mathrsfs}
\usepackage{bbm} 

\usepackage[all,cmtip]{xy}

\usepackage[dvips]{graphicx}

\usepackage[dvips,bookmarks=false]{hyperref} 
\hypersetup{pdfstartview=FitH,pdfhighlight=/O,colorlinks=false}

\newcommand{\mc}[1]{\mathcal{#1}}

\newcommand{\mbb}[1]{\mathbb{#1}}

\newcommand{\p}{\partial}

\newcommand{\Tr}{{\rm Tr}}

\newcommand{\mdots}{,.\,.\,,}

\begin{document}
\title{\large Spinfoams in the holomorphic representation}
\author{Eugenio Bianchi$^{a}$, Elena Magliaro$^{a}$, Claudio Perini$^{ab}$}
\affiliation{${}^a$Centre de Physique Th\'eorique de Luminy\footnote{Unit\'e mixte de recherche (UMR 6207) du CNRS et des Universit\'es de Provence (Aix-Marseille I), de la M\'editerran\'ee (Aix-Marseille II) et du Sud (Toulon-Var); laboratoire affili\'e \`a la FRUMAM (FR 2291).}\;, Case 907, F-13288 Marseille, EU}
\affiliation{${}^b$Dipartimento di Matematica, Universit\`a degli Studi Roma Tre, I-00146 Roma, EU}

\date{April 25, 2010}

\begin{abstract}
We study a holomorphic representation for spinfoams. The representation is obtained via the Ashtekar-Lewandowski-Marolf-Mour\~ao-Thiemann coherent state transform. We derive the expression of the 4d spinfoam vertex for Euclidean and for Lorentzian gravity in the holomorphic representation. The advantage of this representation rests on the fact that the variables used have a clear interpretation in terms of a classical intrinsic and extrinsic geometry of space. We show how the peakedness on the extrinsic geometry selects a single exponential of the Regge action in the semiclassical large-scale asymptotics of the spinfoam vertex. 
\end{abstract}


\maketitle

\section*{Introduction}
Spinfoams \cite{Reisenberger:1996pu,Perez:2003vx} provide a covariant formulation of the dynamics of Loop Quantum Gravity (LQG) \cite{Rovelli:2004tv,Ashtekar:2004eh,Thiemann:book2007} (see \cite{Rovelli:2010wq} for an up-to-date status report). The dynamics is described in terms of transition amplitudes between states belonging to the kinematical Hilbert space of LQG. In particular, when the spin-network basis is used, the transition amplitude is a function of spins and intertwiners associated to the links and the nodes of the spin-network graph. In the recent literature \cite{Livine:2007vk,Engle:2007wy,Freidel:2007py,Kaminski:2009fm}, spinfoams are generally defined in terms of these variables: a space-time configuration in the spinfoam sum is understood as a $2$-complex with faces labeled by $SU(2)$-spins and edges labeled by $SU(2)$-intertwiners.

Different representations of the Hilbert space of LQG can be lifted to the covariant level in spinfoams. In this paper we consider two representations for spinfoams. The first is a \emph{holonomy representation}. This representation allows to write spinfoams into a form closer to a Feynman path-integral: as an integral over the Ashtekar connection smeared along links that slice faces of the spinfoam $2$-complex. The second is the \emph{holomorphic representation}: it is based on the Segal-Bargmann transform for theories of connections introduced by Ashtekar, Lewandowski, Marolf, Mour\~ao and Thiemann \cite{Ashtekar:1994nx}.

Clearly, spinfoams in different representations have the same physical content. What the new representations provide is new insights and new calculational tools. In particular, the holomorphic representation offers a new way of understanding the semiclassical behavior of spinfoams. The holomorphic representation is associated to coherent spin-network states \cite{Thiemann:2000bw,Sahlmann:2001nv,Thiemann:2002vj,Bahr:2007xa},\cite{Bianchi:2009ky}. These states are peaked on a classical intrinsic \emph{and} extrinsic discrete geometry of space. This fact has a remarkable consequence that we discuss below. 

Recently, the large scale asymptotics of the $4d$ spinfoam vertex has been derived \cite{Barrett:2009gg,Barrett:2009mw}. The analysis has been done using the representation in terms of spins and normals introduced by Livine and Speziale \cite{Livine:2007vk}. For a geometric set of boundary data, the asymptotics of the Lorentzian vertex features a sum over two classical solutions. This leads to a cosine of Regge action, similarly to what happens in $3$d for the Ponzano-Regge model. We argue that the presence of the second undesired classical solution in the semiclassical expansion is an artifact of the representation used: the representation in terms of spins and normals diagonalizes the area operator; as a result, it has a maximal spread on its conjugate momentum and cannot completely identify a point in phase space. On the other hand, the holomorphic representation discussed in this paper is based on coherent spin-networks. These states are peaked both on the area and on its conjugate variable, an extrinsic angle. This is enough to select one of the two classical solutions in the semiclassical expansion.

The paper is organized as follows. In section I we discuss the holonomy representation of spinfoams and its relation with the two representations mostly used in the literature: the `spin and intertwiner' representation and the `spin and normals' representation. Then, in section II, we introduce the holomorphic representation as the Segal-Bargmann transform of the holonomy representation. Moreover we discuss the interpretation of the complex variables used in terms of discrete classical geometries. In sections III and IV, we derive the expression of the Euclidean and the Lorentzian spinfoam vertex both in the holonomy and in the holomorphic representation. In section V, we discuss an application of the derived formulae: we show how the peakedness on the extrinsic geometry allows to select a single classical solution in the analysis of the large scale asymptotics of the Lorentzian spinfoam vertex.  

\section*{I. Spin Foams in various representations}
In LQG, the Hilbert space associated to a graph $\Gamma$ embedded in a 3-dimensional hypersurface $\Sigma$ is $\mc{H}_\Gamma=L^2(SU(2)^L/SU(2)^N)$ where $L$ is the number of links of the graph and $N$ the number of nodes. In the `holonomy representation', a state is a gauge invariant function of $SU(2)$ group elements $h_l\;$ ($l=1\mdots L$) that is invariant under $SU(2)$ gauge transformations at nodes, 
\begin{equation}
\Psi(h_l)=\Psi(g_{n_l}\, h_l\, g^{-1}_{n'_l}).
\label{eq:gauge inv}
\end{equation}
Here $n_l$ and $n'_l$ are respectively the node that is source/target of the link $l$. The configuration variables $h_l$ are interpreted as holonomies of the Ashtekar-Barbero connection 
\begin{align}
A_a^i=\Gamma_a^i+\gamma K_a^i
\end{align}
along the link $l$ of the graph ($\Gamma_a^i$ is the spin-connection, $K_a^i$ the extrinsic curvature of the hypersurface $\Sigma$, and the real number $\gamma\neq 0$ is the Barbero-Immirzi parameter).

\begin{figure}[h]
\centering
\includegraphics[width=4cm]{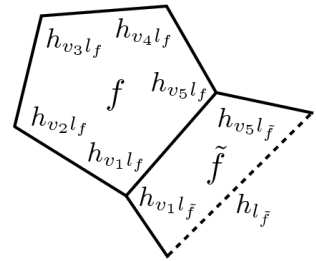}
\caption{The bulk holonomies $h_{vl}$ are associated to the internal face $f$ as well as to the external face $\tilde f$. The notation is the following: $v$ are spinfoam vertices and $l$ are links intersection of the faces with a small 3-sphere centered on the corresponding vertex. The boundary holonomy $h_{\tilde l}$ is associated to the external side (dashed line) of the boundary face $\tilde f$.}
\label{figura}
\end{figure}

Spinfoams provide the transition amplitude from an `in' state to an `out' state: they are maps that correspond to the formal expression
\begin{equation}
W[g^{\text{\tiny $(3)$}}_{\text{in}},g^{\text{\tiny $(3)$}}_{\text{out}}]=\int_{g^{\text{\tiny $(3)$}}_{\text{in}}}^{g^{\text{\tiny $(3)$}}_{\text{out}}} D g^{\text{\tiny $(4)$}}\; \exp i S[g^{\text{\tiny $(4)$}}].
\end{equation}
for the transition amplitude of $3$-geometries in terms of a sum over $4$-geometries \cite{Misner:1957wq,Hawking:1979}. The formalism admits a generalization to `boundary amplitudes' \cite{Rovelli:2004tv,Oeckl:2003vu}. Let $\mc{H}_\Gamma$ be the Hilbert space associated to a graph on the boundary of a $4$-dimensional ball. The boundary amplitude of a state $\Psi\in \mc{H}_\Gamma$ is given by 
\begin{equation}
\langle W | \Psi \rangle=\int \prod_l dh_l\;W(h_l)\; \Psi(h_l),
\label{eq:WPsi}
\end{equation}
where $W(h_l)$ is the spinfoam model in the holonomy representation. The quantity $W(h_l)$ is \emph{local} in space-time, i.e. it is given by a product of elementary vertex-amplitudes $W_v(h_{vl})$ integrated over bulk variables $h_{vl}$:
\begin{equation}
W(h_l)=\sum_{\sigma} \int\! dh^{\text{bulk}}_{vl}\prod_{v\subset \sigma} W_v(h_{vl})\;
\prod_{f\subset\sigma} \delta(\prod_{v\in \p f} h_{vl}).
\label{eq:bulk}
\end{equation}
Here, the sum is over $2$-complexes $\sigma$ with boundary given by the graph $\Gamma$\footnote{The sum over 2-complexes can be generated by an auxiliary Group Field Theory \cite{Oriti:2009wn}. It is usually not well-defined (divergent), and requires suitable gauge-fixing or regularization. For a fixed 2-complex, divergencies are associated to ``bubbles''. In topological (i.e. unphysical) theories, topological invariance implies that those ``bubbles'' can be removed up to a divergent overall factor that depends only on the cutoff. In quantum general relativity the situation is different: ``bubble'' divergencies are true radiative corrections which carry information about the infrared behavior of the theory \cite{Perez:2000fs,Perini:2008pd,Freidel:2009hd,Magnen:2009at,Geloun:2010vj,Krajewski:2010yq,Bonzom:2010ar,Geloun:2010nw}.}. A ``face amplitude'' is present: it is given by a delta function of a product of the holonomies $h_{vl}$ bounding a face $f$ of the $2$-complex. It is needed in order to guarantee the composition law of the spinfoam amplitude \cite{Rov:composition,Bianchi:2010fj}. Notice that we are associating a boundary graph to each vertex of the $2$-complex, therefore the holonomies $h_{vl}$ have a vertex label $v$ and a link label $l$. The association of variables is depicted in Fig.\ref{figura}. deThe construction is the following: we consider a $4$-ball that contains a single vertex of the $2$-complex $\sigma$; the boundary graph associated to the vertex is defined as the intersection of the portion of the $2$-complex contained in the $4$-ball and the boundary of the $4$-ball (a 3-sphere). In particular, the boundary graph has as many links as faces of the $2$-complex intersecting at the vertex. The bulk variables $h_{vl}^{\text{bulk}}$ are $SU(2)$ holonomies associated to these links.
As an example, the vertex amplitude of the Ponzano-Regge model for 3d gravity is given by
\begin{equation}
W_v^{PR}(h_l)=\int \prod_{n=1}^4d g_n\; \prod_{l=1}^6 \delta(g_{n_l}\,h_l\, g^{-1}_{n'_l}).
\label{eq:PR h}
\end{equation}
In the case of the Ponzano-Regge model, the boundary graph is a complete graph with four (3-valent) nodes. Similarly, the vertex proposed by Engle-Pereira-Rovelli-Livine and Freidel-Krasnov (EPRL-FK) \cite{Engle:2007wy,Freidel:2007py} was originally defined on a complete graph with $5$ (4-valent) nodes. More recently, Kaminski-Kisielowski-Lewandowski have proposed a version of the EPRL-FK vertex that provides a generalization to arbitrary boundary graphs \cite{Kaminski:2009fm}.

Above we have defined spinfoams directly in the holonomy representation. In the literature, spinfoam vertices are generally presented using a different representation: the `spin and intertwiner' representation, $W_v(j_l,i_n)$. This representation is related to the holonomy representation by the Peter-Weyl transform (see table \ref{tab}). This is the representation associated to an orthonormal basis of the boundary Hilbert space $\mc{H}_\Gamma$. Let $\Psi_{j_l,i_n}(h_l)$ be a spin-network state,
\begin{equation}
\Psi_{j_l,i_n}(h_l)=\Big(\bigotimes_n i_n\Big)\cdot \Big(\bigotimes_l D^{(j_l)}(h_l)\Big).
\label{eq:spin-network}
\end{equation}
Here $D^{(j)}$ are $SU(2)$ representation matrices and $i_n$ intertwining tensors. The spinfoam vertex in the `spin and intertwiner' representation is given by the amplitude of an element of the spin-network basis
\begin{equation}
W_v(j_l,i_n)=\langle W_v|\Psi_{j_l,i_n}\rangle.
\label{eq:spin and intertwiner}
\end{equation}
For instance, in this representation, the Ponzano-Regge vertex is simply given by a Racah $6j$-symbol \cite{PR:1968},
\begin{equation}
W^{PR}_v(j_l)=\textstyle{\frac{1}{\sqrt{\prod_l (2j_l+1)}}}\{6j\}.
\end{equation}
In the definition and in the analysis of the asymptotics of the new spinfoam vertices, a third representation has turned out to be useful \cite{Conrady:2008ea,Conrady:2008mk,Barrett:2009gg,Barrett:2009mw,Dowdall:2009eg}. It is a representation in terms of a spin and two unit-normals per link of the graph.  This representation is associated to spin-networks with nodes labeled by Livine-Speziale coherent intertwiners \cite{Livine:2007vk,Conrady:2009px}. A coherent intertwiner in $\text{Inv}(\otimes_{e=1}^E H^{(j_e)})$ is labeled by $E$ unit-vectors $\vec{n}_e$ satisfying the closure condition $\sum_e j_e \vec{n}_e =0$. They are defined in terms of Bloch $SU(2)$-coherent states\footnote{Let $R_{\vec{n}}$ be the matrix that rotates the vector $\vec{e}_3=(0,0,1)$ into the unit-vector $\vec{n}$, leaving the vector $\vec{e}_3\times\vec{n}$ invariant; we call $h_{\vec{n}}$ the associated $SU(2)$ element, $h_{\vec{n}}=D^{(1/2)}(R_{\vec{n}})$. A Bloch coherent state in direction $\vec{n}$ is obtained rotating the maximum-eigenvalue eigenstate of the spin in the direction $\vec{e}_3$ to the direction $\vec{n}$, i.e. $|j,\vec{n}\rangle=D^{(j)}(h_{\vec{n}})|j,+j\rangle$. These are the states used in the description of the magnetic moment of a nucleus in an external magnetic field \cite{Bloch:1946zza}.\label{fn:Bloch}} $|j_e,\vec{n}_e\rangle$  and are given by the following formula
\begin{equation}
\Phi(\vec{n}_e)=\int_{SU(2)}\hspace{-1em}dg\; \bigotimes_{e=1}^{E} D^{(j_e)}(g)|j_e,\vec{n}_e\rangle.
\end{equation}
A spin-network state with coherent intertwiners at nodes is given by
\begin{equation}
\Psi_{j_l,\vec{n}_l,\vec{n}'_l}(h_l)=\Big(\bigotimes_n \Phi_n(\vec{n}_l)\Big) \cdot \Big(\bigotimes_l D^{(j_l)}(h_l)\Big).
\label{eq:LS spin-network}
\end{equation}
These states form an overcomplete basis of the Hilbert space $\mc{H}_\Gamma$ with the measure on normals derived in \cite{Conrady:2009px}. 

A spinfoam vertex can be written in the `spin and normals' representation,
\begin{equation}
W_v(j_l,\vec{n}_l,\vec{n}'_l)=\langle W_v|\Psi_{j_l,\vec{n}_l,\vec{n}'_l}\rangle.
\end{equation}
The reason why this representation is so useful in the semiclassical analysis of the vertex is that the normals $\vec{n}$ are classical variables, as opposed to the intertwiners in (\ref{eq:spin and intertwiner}) that are quantum numbers. This is in fact the original motivation of \cite{Livine:2007vk} for introducing coherent intertwiners. On the other hand, the spins $j_l$ are still quantum numbers\footnote{Quantum numbers refer generally to discrete eigenvalues of observables, as in the case of geometric operators (length, area and volume operator) in Loop Quantum Gravity.}. In the following we introduce a new representation where the classical counterparts of the spins and of their conjugate momenta appear.
\begin{table}[t]%
{\large\begin{equation*}
\xymatrix{
\;W(h_l)\; \ar[d]_{PW} \ar[r]^{SB} &\;W(H_l)\;\ar@{.>}[d]\\
\;W(j_l,i_n)\; \ar[r]^{LS} &\;W(j_l,\vec{n}_l,\vec{n}'_l)\;}
\end{equation*}}
\caption{The various representations of spinfoams are related by transforms as shown in the diagram. The Peter-Weyl transform (PW) sends the holonomy representation $W(h_l)$ in the `spin and intertwiner' representation $W(j_l,i_n)$. The `spin and normals' representation $W(j_l,\vec{n}_l,\vec{n}'_l)$ is obtained from the `spin and intertwiner' representation via the Livine-Speziale transform (LS). The topic of this paper is the holomorphic representation $W(H_l)$. It is obtained from the holonomy representation via the Segal-Bargmann transform (SB). The expression for the 4d vertex amplitudes in the holomorphic representation is eq. (\ref{eq:WEPRL H}) for the Euclidean case, and eq. (\ref{eq:Lor H}) for the Lorentzian case. In the semiclassical analysis of spinfoams, there is an interesting interplay of the holomorphic and the `spin and normals' representations. This is discussed in section V.
}
\label{tab}
\end{table}

\section*{II. The holomorphic representation}
In this section we introduce a new representation of spinfoams, a representation based on \emph{coherent spin-networks} \cite{Bianchi:2009ky},\cite{Thiemann:2000bw,Sahlmann:2001nv,Thiemann:2002vj,Bahr:2007xa}. 
Coherent spin-network states provide an overcomplete basis of the boundary Hilbert space $\mc{H}_\Gamma$. They are labeled by an element $H_l$ of $SL(2,\mbb{C})$ per link of the graph and are defined by
\begin{equation}
\Psi_{H_l}(h_l)=\int \prod_n dg_n\;\prod_l K_{t_l}(g_{n_l}\,h_l\, g^{-1}_{n'_l}\, H_l^{-1}).
\label{eq:coherent spin-networks}
\end{equation}
Here, $t_l$ are positive real numbers and $K_t$ is the analytic continuation to $SL(2,\mbb{C})$ of the heat-kernel on $SU(2)$,
\begin{equation}
K_t(h)=\sum_j (2j+1) e^{-j(j+1)\frac{t}{2}}\; \Tr D^{(j)}(h).
\end{equation}
As shown in \cite{Ashtekar:1994nx}, coherent spin-networks provide a Segal-Bargmann transform for Loop Quantum Gravity. 
Given a state $\Psi_f(h_l)$, its scalar product with a coherent spin-network $\Psi_{H_l}(h_l)$ defines a function $\Phi_f(H_l)$ that is holomorphic in $H_l$,
\begin{equation}
\Phi_f(H_l)=\langle\Psi_f|\Psi_{H_l}\rangle=\int\prod_l dh_l\;\overline{\Psi_f(h_l)}\;\Psi_{H_l}(h_l),
\label{eq:trasform}
\end{equation}
and belongs to the Hilbert space 
\begin{equation}
\mc{H}L^2(SL(2,\mbb{C})^L,d\nu_t(H)^L)
\end{equation}
 of holomorphic functions normalizable with respect to a measure $d\nu_t(H)^L$ (see \cite{Ashtekar:1994nx} and the appendix of \cite{Bianchi:2009ky} for details\footnote{We have $d\nu_t(H)=\Omega_{2t}(H)dH$, where $dH$ is the Haar measure on $SL(2,\mathbb C)$, and $\Omega_t(H)$ is just the heat-kernel on $SL(2,\mathbb C)/SU(2)$, regarded as a $SU(2)$-invariant function on $SL(2,\mathbb C)$.}). This result is based on the seminal work of Hall \cite{Hall1994}. Moreover the $SL(2,\mbb{C})$ labels $H_l$ provide a parametrization of the phase space of the theory as captured by a graph. The peakedness properties of these states and their geometrical interpretation in terms of classical holonomies and fluxes is well-studied within Loop Quantum Gravity \cite{Thiemann:2000bw,Sahlmann:2001nv,Thiemann:2002vj,Bahr:2007xa}. However their use in spinfoams has remained largely unexplored until recently \cite{Bianchi:2009ky}. In the following we show how coherent spin-networks and the associated holomorphic representation can be fruitfully used in the spinfoam setting.

A spinfoam vertex can be expressed in the holomorphic representation using formula (\ref{eq:WPsi}),
\begin{equation}
W_v(H_l)=\langle W_v|\Psi_{H_l}\rangle.
\end{equation}
As an example, the Ponzano-Regge vertex amplitude (\ref{eq:PR h}) in the holomorphic representation is given by the following formula:
\begin{equation}
W_v^{PR}(H_l)=\int \prod_n dg_n\;\prod_l K_{t_l}(\, (g_{n_l}\,H_l\, g^{-1}_{n'_l})^{-1}).
\end{equation} 
The advantage of using the holomorphic representation is that the $SL(2,\mbb{C})$ labels admit a clear interpretation in terms of classical discrete geometries. Here we discuss two equivalent descriptions of the discrete geometry associated to the $SL(2,\mbb{C})$ variables: the first (a) is the one proper of canonical LQG and is due to Sahlmann-Thiemann-Winkler \cite{Thiemann:2000bw,Sahlmann:2001nv,Thiemann:2002vj}, the second (b) is the one mostly used in the covariant spinfoam setting, it is in terms of Freidel-Speziale variables for twisted geometries \cite{Freidel:2010aq} and has been introduced by the authors in \cite{Bianchi:2009ky}.

\vspace{0.5em}
(a) Let $\Sigma$ be the boundary $3$-manifold and choose a classical configuration of the Ashtekar connection $A$ and its conjugate momentum, the electric field $E$. Moreover, let $\Delta_\Sigma$ be a cellular decomposition of $\Sigma$ and $\Gamma$ a graph dual to this decomposition. The flux of the electric field through a face $f_l$ of $\Delta_\Sigma$ determines a $su(2)$ algebra element $E_l=\int_{f_l} E$. The holonomy of the connection along a link $l$ of the graph $\Gamma$ determines a $SU(2)$ group element $h_l=P\exp\int_l A$. The set of couples $(h_l,E_l)$, one per link of the graph, determines a point in a truncation of the phase space of General Relativity as captured by the graph $\Gamma$. The couple $(h_l,E_l)$ can be written as an element of the complexification of $SU(2)$ using the polar decomposition\footnote{In formula \eqref{eq:complexification}, the factor $\frac{t}{8\pi G\hbar\gamma}$ in front of $E_l$ is required in order to have the correct interpretation of the $SL(2,\mathbb C)$ variables in terms of classical fluxes and holonomies, as a careful analysis of the expectation values of geometric operators elucidate. On the expecation values of elementary operators, see \cite{Bianchi:2009ky}.} of $SL(2,\mbb{C})$:
\begin{equation}
H_l=\;h_l\;\exp(i\frac{E_l}{8\pi G \hbar \gamma}t_l).
\label{eq:complexification}
\end{equation}
This is the Cartan decomposition of a Lorentz transformation into an $SU(2)$ rotation and a boost. Coherent spin-networks with labels as in (\ref{eq:complexification}) are peaked on the classical configuration $(h_l,E_l)$.

\vspace{0.5em}
(b) Let $\Delta_\Sigma$ be a cellular decomposition of the boundary $3$-manifold $\Sigma$ (the cells of the decomposition can be simplicial or even polyhedral). Equip separately each cell with a Euclidean $3$-geometry. As a result, the shape of each cell is fixed and the area and the normals to its faces can be computed. Now we consider two cells that share a face, and require that the area of the face is the same when seen from either of the two cells. Therefore, for each face $f_l$ of $\Delta_\Sigma$, we have as variables its area $\mathscr{A}_l$ and two normals $\vec{n}_l$ and $\vec{n}'_l$. In general the geometry these data determine is twisted in the sense of \cite{Freidel:2010aq}. A special case is the one where shared faces have congruent shapes. In this case, a continuous piecewise-flat geometry is obtained. In the case of a simplicial decomposition, this is a Regge geometry\footnote{On the encoding of simplicial geometry in a lattice gauge theory formulation of quantum gravity, see \cite{Bonzom:2009wm,Oriti:2009wg,Dittrich:2008ar}. For a discussion of the matching of lengths in LQG see sec.5 of \cite{Bianchi:2008es}. For a non-commutative flux representation of LQG and its simplicial interpretation see \cite{Baratin:2010wi}.} \cite{Regge:1961px}.

Besides areas and $3$-normals, there is another quantity of interest. Let us consider two cells sharing a face $f_l$ and think them as embedded in $4$-dimensional Minkowski space. We call $N_l^I$ and ${N'_l}^I$ the $4$-normals to the two cells. Their scalar product determines an angle $\Theta_l$ with the meaning of extrinsic curvature, $\cosh \Theta_l=-\eta_{IJ} N_l^I {N'_l}^J$, where $\eta_{IJ}$ is the Minkowski metric. We call $\xi_l$ the extrinsic angle times the Immirzi parameter\footnote{The parameter $\xi$ codes the rotation generated by the extrinsic curvature term $\gamma K_a^i$ of Ashtekar-Barbero connection.} $\gamma$,
\begin{equation}
\xi_l=\gamma \Theta_l.
\end{equation}
Summarizing, for each link of the graph $\Gamma$ dual to $\Delta_\Sigma$, we have the following set of variables: 

\vspace{.5em}

- an area $a_l=\frac{\mathscr{A}_l}{8\pi G \hbar \gamma}$,

\vspace{.5em}

- an \emph{extrinsic} angle $\xi_l$, 

\vspace{.5em}

- two unit-normals $\vec{n}_l$ and $\vec{n}'_l$. 

\vspace{.5em}

\noindent Out of these variables we can build a $SL(2,\mbb{C})$ element $H_l$ given by the following expression
\begin{equation}
H_l=h_{\vec{n}_l}\,e^{-i z_l \frac{\sigma_3}{2}}\,h^{-1}_{-\vec{n}'_l},
\label{eq:Hl}
\end{equation}
where the complex number $z_l$ is given in terms of $\xi_l$ and $a_l$ by
\begin{equation}
z_l=\xi_l+i a_l t_l,
\label{eq:zl}
\end{equation}
and $h_{\vec{n}}$ is the $SU(2)$ element associated to a vector $\vec{n}$ as explained in the footnote${}^{\text{\ref{fn:Bloch}}}$. The reason for this parametrization is that coherent spin-networks (\ref{eq:coherent spin-networks}) with $SL(2,\mbb{C})$ labels specified as in (\ref{eq:Hl}), for large area $a_l$ reduce to superpositions over spins of Livine-Speziale spin-networks (\ref{eq:LS spin-network}) \cite{Bianchi:2009ky}. Moreover, the superposition over spins is a Gaussian centered in $j_l\approx a_l$ times a phase term $\exp(-i \xi_l j_l)$. This is the superposition originally proposed by Rovelli in the analysis of the graviton propagator \cite{Rovelli:2005yj}. The interpretation of the $SL(2,\mbb{C})$ labels in terms of twisted geometries will be important in section V where we discuss the semiclassical behavior of spinfoams.

The relation between the description in terms of fluxes and holonomies (a) and the description in terms of areas, angles and normals (b) can be easily derived. Writing (\ref{eq:Hl}) as the polar decomposition
\begin{equation}
H_l=\big(h_{\vec{n}_l}\,e^{-i \gamma \Theta_l \frac{\sigma_3}{2}}\,h^{-1}_{-\vec{n}'_l}\big)\; \exp\big( i \frac{\mathscr{A}_l}{8\pi G \hbar \gamma}\vec{n}'_l\cdot\frac{i\vec{\sigma}}{2}\,t_l\big)
\end{equation}
we recognize the variables used in (\ref{eq:complexification}),
\begin{align}
h_l=&\,P\exp\int_l A =h_{\vec{n}_l}\,e^{-i \gamma \Theta_l \frac{\sigma_3}{2}}\,h^{-1}_{-\vec{n}'_l},\\
E_l=&\int_{f_l}E=\mathscr{A}_l \,\vec{n}'_l\cdot\frac{i\vec{\sigma}}{2}.
\end{align}

\section*{III. 4d Euclidean vertex amplitude in the holomorphic representation}\label{sec:euclidean}
The 4d spinfoam vertex amplitude \cite{Engle:2007wy,Freidel:2007py} is generally defined in terms of its components on a spin-network basis. In the Euclidean case, for Immirzi parameter $0<\gamma<1$, it is given by 
\begin{equation}
W_v(j_l, i_n)=\sum_{i_n^+,i_n^-} \{15J\}(j^+_l, i^+_n; j^-_l, i^-_n)\;\prod_n f_{i_n^+,i_n^-}^{i_n}
\label{eq:EPRL j i}
\end{equation}
with $j^\pm_l=\frac{1\pm\gamma}{2}j_l$, $\{15J\}$ the $15j$-symbol for $SU(2)\times SU(2)$ and the fusion coefficients $f_{i_n^+,i_n^-}^{i_n}$ defined in \cite{Engle:2007wy} (see also \cite{Alesci:2008un}). This representation has been used in numerical investigations \cite{Magliaro:2007nc}. Contracting it with the spin-network basis (\ref{eq:spin-network}) leads to a rather simple integral formula for the vertex in the holonomy representation.

Let $Y$ be a map from the representation $H^{(j)}$ of $SU(2)$ to the representation $H^{(j^+,j^-)}$ of the Euclidean group $Spin(4)\simeq SU(2)\times SU(2)$. The map $Y$ is such that it commutes with the diagonal action of $SU(2)$ on $SU(2)\times SU(2)$. In the canonical basis, the map $Y$ is given by Clebsh-Gordan coefficients  
\begin{equation}
Y_m^{m^+,m^-}=\langle j^+,m^+;j^-,m^-|j,m\rangle.
\end{equation}
Notice that $j^+ + j^-=j$, therefore the map $Y$ corresponds to a projection to the highest weight representation in the tensor product $H^{(j^+)}\otimes H^{(j^-)}$. Using this map, the vertex amplitude in the holonomy representation can be written as
\begin{equation}
W_v(h_l)=\int_{Spin(4)^5}\prod_n dG_n\; \prod_l P(h_l,G_{n_l} G_{n'_l}^{-1}).
\label{eq:W hl}
\end{equation}
The integrals on $Spin(4)$ impose the Euclidean invariance of the vertex, and the integrand is given by a product of `face' terms $P$ given by the following distribution over $L^2(SU(2))\otimes L^2(Spin(4))\simeq L^2(SU(2)^3)$:
\begin{equation}
P(h,G)=\sum_{j} (2j+1) \text{Tr}\Big(D^{(j)}(h)\; Y^\dagger D^{(j^+\!,j^-)}(G^{-1})Y\Big).
\end{equation}
Expression (\ref{eq:W hl}) can be easily put in the holomorphic representation using the Segal-Bargamann transform \cite{Ashtekar:1994nx}:
\begin{equation}
W(H_l)=\int_{Spin(4)^5}\prod_n dG_n\; \prod_l P_{t_l}(H_l,G_{n_l} G_{n'_l}^{-1})
\label{eq:WEPRL H}
\end{equation}
where
\begin{align}
P_t(H,G)=&\sum_{j} (2j+1)\; e^{-j(j+1)\frac{t}{2}}\;\times\label{eq:P H}\\
&\times\; \text{Tr}\Big(D^{(j)}(H)\; Y^\dagger D^{(j^+\!,j^-)}(G^{-1})Y\Big).\nonumber
\end{align}
Notice that the last expression can be seen as a distribution on $\mathcal HL^2(SL(2,\mathbb C))\otimes L^2(Spin(4))$.
\section*{IV. 4d Lorentzian vertex amplitude in the holomorphic representation}\label{sec:lorentzian}
In this section we discuss the definition of the 4d Lorentzian vertex \cite{Pereira:2007nh,Engle:2007wy} and derive its holonomy and its holomorphic representation.

Let $\Psi(h_l)$ be a gauge invariant state on a graph bounding the spinfoam vertex
\begin{equation}
\Psi(h_l)=\sum_{j_l} \Psi^{(j_l)}\cdot \big(\bigotimes_l D^{(j_l)}(h_l)\big).
\label{eq:bstate}
\end{equation}
The 4d Lorentzian vertex amplitude for this state is obtained in two steps:
\begin{itemize}
  \item[(i)] we map the $SU(2)$-boundary state $\Psi(h_l)$ into a $SL(2,\mbb{C})$-boundary state $(f\circ \Psi)(G_l)$ using a ($\gamma$ dependent) map $f$ defined below;
  \item[(ii)] we impose Lorentz invariance at nodes.
\end{itemize} 
This provides a compact definition of the vertex amplitude as\footnote{Notice that one of the integrals over $SL(2,\mbb{C})$ has to be gauge-fixed in order to avoid having a divergent expression \cite{Engle:2008ev}.}
\begin{equation}
\langle W_v|\Psi\rangle=\int_{SL(2,\mbb{C})^N}\prod_n dG_n\;f \Psi(G_{n_l} G^{-1}_{n'_l}).
\label{eq:WvPsi}
\end{equation}
The EPRL-FK vertex corresponds to a specific choice of $f$ defined as follows. Let us consider the principal series of unitary irreducible representations $(\rho,n)$ of $SL(2,\mbb{C})$ \cite{Ruhl} with Casimir operators given by 
\begin{align}
&C_1=\frac{1}{2}(n^2-\rho^2-4),\\
&C_2=n\rho.
\end{align}
They decompose into irreducible representations $j$ of a subgroup $SU(2)$ as
\begin{equation}
H_{SL(2,\mbb{C})}^{(\rho,n)}=\bigoplus_{j\geq\frac{n}{2}}H_{SU(2)}^{(j)}.
\end{equation}
Consider the representation with labels $\rho=2\gamma j$, $n=2 j$ and let $Y$ be the embedding of $H_{SU(2)}^{(j)}$ into the lowest weight representation of $H_{SL(2,\mbb{C})}^{(\gamma 2j,2j)}$. Calling $|(\gamma 2j,2j); j,m\rangle$ its canonical basis, the map $Y$ is given by
\begin{equation}
Y=\sum_{m=-j}^{+j}|(\gamma 2j,2j); j,m\rangle\langle j,m|.
\end{equation}
Now we define the map 
\begin{equation}
f:L^2(SU(2)^L/SU(2)^N)\to L^2(SL(2,\mbb{C})^L)
\end{equation}
 in terms of $Y$ as follows. Given the state (\ref{eq:bstate}), we have
\begin{equation}
f \Psi(G_l)=\sum_{j_l} \Psi^{(j_l)}\cdot \big(\bigotimes_l Y^\dagger D^{(\gamma 2j_l,2j_l)}(G_l) Y\big).
\label{eq:f}
\end{equation}
The fact that $f\Psi$ is square-integrable follows easily from the ortogonality of the Wigner matrices in the principal series (on the definition and finiteness properties of $SL(2,\mathbb C)$ spin-networks, see \cite{Engle:2008ev,Livine:2002ak}).
Using the definition (\ref{eq:WvPsi}),(\ref{eq:f}) of the Lorentzian vertex, it is easy to derive its holonomy representation. Let $\Psi_{h'_l}(h_l)$ be the $SU(2)$ gauge invariant delta function on the boundary graph
\begin{equation}
\Psi_{h'_l}(h_l)=\int_{SU(2)^N}\prod_n dg_n \prod_l \delta(g_{n_l}\,h_l g_{n'_l}^{-1}\,{h'}_l^{-1}).
\label{eq:delta}
\end{equation}
Its vertex amplitude provides the holonomy representation of the Lorentzian vertex. It is given by the following integral expression
\begin{equation}
W_v(h_l)=\langle W_v|\Psi_{h_l}\rangle=\int\prod_n dG_n\;\prod_l P(h_l,G_{n_l}G^{-1}_{n'_l}),
\label{eq:Lor h}
\end{equation}
where the integral is over $SL(2,\mbb{C})^N$ and the `face' term is
\begin{equation}
P(h,G)=\sum_{j} (2j+1) \text{Tr}\Big(D^{(j)}(h)\; Y^\dagger D^{(\gamma 2j,2j)}(G^{-1})Y\Big)
\end{equation}
similarly to what happens in the Euclidean case. Its Segal-Bargmann transform corresponds to the vertex amplitude of a coherent spin-network. It is given by
\begin{equation}
W_v(H_l)=\langle W_v|\Psi_{H_l}\rangle=\int\prod_n dG_n\;\prod_l P_{t_l}(H_l,G_{n_l}G^{-1}_{n'_l}),
\label{eq:Lor H}
\end{equation}
where the `face' term now is
\begin{align}\label{faceterm}
P_t(h,G)=&\sum_{j} (2j+1) e^{-j(j+1)\frac{t}{2}} \times\\
&\times\text{Tr}\Big(D^{(j)}(H)\; Y^\dagger D^{(\gamma 2j,2j)}(G^{-1})Y\Big).\nonumber
\end{align}
We report also the `spin and normals' representation of the Lorentzian vertex
\begin{align}
&W_v(j_l,\vec{n}_{l},\vec{n}'_l)\equiv\;\langle W_v|\Psi_{j_l,\vec{n}_{l},\vec{n}'_l}\rangle= \label{eq:Lor normals}\\
&=\int\prod_n dG_n\;\prod_l \langle j_l, -\vec{n}_l|Y^\dagger D^{(\gamma 2j,2j)}(G_{n_l}G^{-1}_{n'_l})Y |j_l,\vec{n}'_l\rangle.
\nonumber
\end{align}
This expression is the one used in the analysis of the asymptotics of the vertex \cite{Barrett:2009mw}. In next section we uncover an interesting interplay of the representations (\ref{eq:Lor H}) and (\ref{eq:Lor normals}) of the spinfoam vertex in the analysis of its semiclassical behavior.

\section*{V. Semiclassical analysis: role of the extrinsic curvature in the large spin asymptotics}
The holomorphic representation is associated to a set of states that have semiclassical properties. This fact makes the holomorphic representation the appropriate tool for studying the semiclassical behavior of spinfoams. We illustrate this point using a simple quantum mechanical system, and then discuss it in the case of the 4d Lorentzian vertex amplitude.

Let us consider a particle on a line. The evolution operator in the position representation is given by
\begin{equation}
W(x_1,x_2)=\langle x_2|e^{-\frac{i}{\hbar}HT}|x_1\rangle\\
=\int_{x_1}^{x_2}\hspace{-.5em}D[x(t)]\;e^{\frac{i}{\hbar}S[x(t)]},
\label{eq:W(x_1,x_2)}
\end{equation}
where $H$ is the Hamiltonian operator for the particle, $T$ the evolution time, and $S[x(t)]$ the classical action. In the formal semiclassical limit $\hbar\to 0$, the path integral in (\ref{eq:W(x_1,x_2)}) is dominated by classical solutions $x_{(k)}(t)$ starting at $x_1$ and ending at $x_2$. There can be more than a single classical solution and the stationary phase approximation  of the path integral is given by a sum over all classical contributions \cite{FeynHibbs:1965}
\begin{equation}
W(x_1,x_2)\stackrel{\hbar\to 0}{\approx} \sum_k A_{\text{\tiny $(k)$}}(x_1,x_2)\; e^{\frac{i}{\hbar} S_{\text{\tiny $(k)$}}(x_1, x_2)}.
\label{eq:h to 0}
\end{equation}
Here $S_{\text{\tiny $(k)$}}(x_1, x_2)=S[x_{\text{\tiny $(k)$}}(t)]$ is the Hamilton function, that is the action evaluated on one of the classical solutions. $A_{\text{\tiny $(k)$}}(x_1, x_2)$ is a slowly varying function of $x_1$ and $x_2$. 

Notice that, if we expand the Hamilton function around the points $x_1=q_1$ and $x_2=q_2$, the linear terms in the expansion are proportional to the initial and final momenta. In fact the derivatives of the Hamilton function are
\begin{align}
p_2^{\text{\tiny $(k)$}}(x_1, x_2)&\;=\frac{\p}{\p x_2} S_{\text{\tiny $(k)$}}(x_1, x_2),\label{eq:p}\\
p_1^{\text{\tiny $(k)$}}(x_1, x_2)&\;=-\frac{\p}{\p x_1} S_{\text{\tiny $(k)$}}(x_1, x_2)\nonumber
\end{align}
where $p_1^{\text{\tiny $(k)$}}(x_1, x_2)$ and $p_2^{\text{\tiny $(k)$}}(x_1, x_2)$ are the initial and final momenta of the classical solution $x_{\text{\tiny $(k)$}}(t)$. This fact has two important consequences:
\begin{itemize}
  \item[(i)] Let us consider a Gaussian wave packet peaked at the initial position $q_1$ and initial momentum $p_1$, and study its evolution with the dynamics (\ref{eq:W(x_1,x_2)}). In the $\hbar\to 0$ limit, only one of the terms in the sum over classical solutions in (\ref{eq:h to 0}) contributes to its evolution. Specifically, the term that contributes is the exponential of the Hamilton function for the classical solution $x_{\text{\tiny $(0)$}}(t)$ that has $q_1,p_1$ as initial conditions. 
  \item[(ii)] The transition amplitude to a Gaussian wave packet that is peaked at a final position $q_2$ and final momentum $p_2$ is suppressed unless these are the final position and momentum of the classical solution $x_{\text{\tiny $(0)$}}(t)$ that has $q_1,p_1$ as initial conditions.
\end{itemize}
These two facts make the holomorphic representation especially well-suited for the analysis of the semiclassical behavior of the dynamics.

The holomorphic representation for a particle on a line is associated with coherent states given by the analytic continuation in $x_0$ of the heat-kernel on a line
\begin{equation}
K_t(x,x_0)=\frac{1}{\sqrt{2\pi t}} e^{-\frac{(x-x_0)^2}{2t}}.
\end{equation}
When $x_0$ is continued to the complex number $z_0=q_0+\frac{i}{\hbar}p_0 t$, the state $K_t(x,z)$ is peaked both on the position $q_0$ of the particle \emph{and} on its momentum $p_0$. In fact the state has a phase term $\exp \frac{i}{\hbar} x p_0$ coding the peakedness on the momentum,
\begin{equation}
K_t(x,z_0)=\frac{1}{\sqrt{2\pi t}} e^{-\frac{(x-q_0)^2}{2t}} \; e^{+\frac{i}{\hbar}(x-q_0) p_0}\;e^{+\frac{t}{2\hbar^2}p_0^2}.
\end{equation}
The holomorphic representation of the evolution operator is
\begin{equation}
W(z_1, z_2)=\!\int\!\! dx_1\!\int\!\! dx_2\, \overline{K_t(x_2,z_2)}\,W(x_1, x_2)\,K_t(x_1,z_1).
\label{eq:KWK}
\end{equation}
We call the couple $(z_1,z_2)= (q_1,p_1;q_2,p_2)$ \emph{classical} if it corresponds to the boundary conditions of a classical solution $x_{\text{\tiny $(0)$}}(t)$ of the dynamics. In the $\hbar\to 0$ limit, the transition amplitude $W(z_1,z_2)$ is maximized when the couple $(z_1,z_2)$ corresponds to a classical solution. Otherwise, if $(z_1,z_2)$ is not classical, a rapidly oscillating term $\exp \frac{i}{\hbar} (p_{\text{\tiny $(k)$}}-p_1) x_1$ appears in the integral (\ref{eq:KWK}) with the effect that the transition amplitude $W(z_1,z_2)$ is suppressed. 

To summarize: the feature of the holomorphic representation is that at most one classical trajectory contributes to the semiclassical expansion of the transition amplitude. A similar phenomenon happens in spinfoams.\\

Let us consider the Lorentzian spinfoam vertex (\ref{eq:Lor H}) and focus on the case of a graph $\Gamma$ dual to the boundary of a $4$-simplex as originally done in \cite{Pereira:2007nh,Engle:2007wy}. The amplitude  $W_v(H_l)$ depends on ten elements of $SL(2,\mbb{C})$, one per link of the graph. We can parametrize them in terms of twisted geometries using the variables $(a_l,\xi_l,\vec{n}_l,\vec{n}'_l)$, as explained in section II. The semiclassical limit $\hbar\to 0$ corresponds to the large area asymptotics, $a_l\gg1$. In the following, we discuss the large area asymptotics of $W_v(H_l)$.

Notice that the large $a_l$ asymptotics of $D^{(j)}(e^{a_l \frac{\sigma_3}{2}})$ is proportional to the projector to the highest magnetic number $|j, +j\rangle\langle j, +j|$. As shown in \cite{Bianchi:2009ky}, a consequence of this fact is that the large area asymptotics of $H_l$ in the definition of coherent spin-networks reproduces a Gaussian with a phase on spins, times Bloch coherent states: 
\begin{align}
&e^{-j_l(j_l+1)\frac{t}{2}}D^{(j_l)}(H_l) \;\stackrel{a_l\gg 1}{\approx}\exp(-i\xi_l j_l) \times\\
&\qquad\times \exp\big(-(j_l-a_l)^2\frac{t}{2}+a_l^2\frac{t}{2}\big)\;|j_l,\vec{n}_l\rangle\langle j_l, -\vec{n}'_l|.\quad\nonumber
\end{align}
When gauge invariance at nodes is imposed, Bloch coherent states intertwine and reproduce Livine-Speziale states. As a result, the large area asymptotics of the spinfoam vertex in the holomorphic representation is given by the vertex in the `spin and normals' representation (\ref{eq:Lor normals}), summed over spins with a Gaussian weight times a phase in spins:
\begin{align}
W(H_l)\stackrel{a_l\gg1}{\approx}&\sum_{j_l} \;W(j_l,\vec{n}_l,\vec{n}'_l)\;\exp({-i\sum_l\xi_l j_l})\times\label{eq:W=sumj}\\ &\times\exp(-\sum_l(j_l-a_l)^2\frac{t_l}{2} +\sum_l a_l^2\frac{t}{2}).\nonumber
\end{align}
Now we consider the sum over spins. The Gaussian in the second line of (\ref{eq:W=sumj}) is peaked on a large value of the spins, $j_l\approx a_l\gg1$. As a result, if we want to estimate the sum over spins, it is enough to know the large spin behavior of the quantity $W(j_l,\vec{n}_l,\vec{n}'_l)$. This quantity has been studied in detail by Barrett-Dowdall-Fairbairn-Hellmann-Pereira \cite{Barrett:2009mw}. We briefly recall their results. They compute the large spin asyptotics of the 4d Lorentzian vertex in the spins-normals representation (\ref{eq:Lor normals}). The result found is the following:

\vspace{.5em}

(i) when the spins and the normals identify the geometry of the boundary of a $4$-simplex in 4d Minkowski space, the asymptotics gets two contributions:
\begin{equation}
W(j_l,\vec{n}_l,\vec{n}'_l)\stackrel{j_l\gg 1}{\approx}\sum_{p=\pm 1} A_{(p)}\;\exp( p\, i \sum_l \gamma j_l\; \Theta_l).
\label{eq:Barrett}
\end{equation}
Each contribution features a rapidly oscillating phase term given by the Regge action for discrete gravity \cite{Regge:1961px},
\begin{equation}
S_{\text{R}}(j_l,\vec{n}_l,\vec{n}'_l)=\sum_l \gamma j_l\; \Theta_l(j_l,\vec{n}_l,\vec{n}'_l).
\end{equation}
This is the action for a single $4$-simplex with faces of area $A_l=8\pi G \hbar\,\gamma j_l$, extrinsic angle $\Theta_l$, and a given choice of parity.

\vspace{.5em}

(ii) On the other hand, if the boundary data $(j_l,\vec{n}_l,\vec{n}'_l)$ do not identify the edge lengths of a $4$-simplex but give only a twisted geometry, then the vertex amplitude is exponentially suppressed.
  
The two semiclassical contributions $p=\pm1$ in (\ref{eq:Barrett}) correspond to parity-reversed classical solutions. We argue here that the appearence of a sum over classical solutions in the asymptotics of $W(j_l,\vec{n}_l,\vec{n}'_l)$ is exactly analogous to what happens in the simpler case discussed at the beginning of this section (\ref{eq:h to 0}): it is due to the fact that the `spin and normals' representation does not identify a point in phase space. Let us see now what happens in the holomorphic representation (\ref{eq:W=sumj}) when we use the result (\ref{eq:Barrett}).

Our boundary data now are not just areas and normals $(a_l,\vec{n}_l,\vec{n}'_l)$, there are also the angles $\xi_l$ that are conjugate to the areas. Let us assume that the areas and the normals are compatible with the edge-lengths of a $4$-simplex\footnote{If it is not the case the amplitude is suppressed because $W(j_l,\vec{n}_l,\vec{n}'_l)$ is.}. Plugging (\ref{eq:Barrett}) into (\ref{eq:W=sumj}) we have
\begin{align}
W(H_l)\stackrel{a_l\gg1}{\approx}&\sum_{p=\pm 1} \sum_{j_l}  A_{(p)}\;\exp\big(p\,i  S_R(j_l)- i \sum_l \xi_l j_l\big)\times\nonumber\\\quad&\times\exp(-\sum_l(j_l-a_l)^2\frac{t_l}{2} +\sum_l a_l^2\frac{t}{2}).
\end{align}
In the sum over spins there is now a rapidly oscillating phase term
\begin{equation}
\exp i (p\,\gamma \Theta^{\text{\tiny $(0)$}}_l -\xi_l) j_l.
\end{equation}
Analogously to expression (\ref{eq:p}), the quantity $\Theta^{\text{\tiny $(0)$}}_l$ is the momentum computed out of Regge action for given boundary data:
\begin{equation}
\Theta^{\text{\tiny $(0)$}}_l=\frac{\p S_R}{\p (\gamma j_l)}(a_l)=\Theta_l(a_l).
\end{equation}
Therefore the sum over spins is suppressed unless the variable $\xi_l$ in the boundary data is chosen to coincide with $+\gamma\Theta^{\text{\tiny $(0)$}}_l$ or $-\gamma\Theta^{\text{\tiny $(0)$}}_l$. In either case, only one of the two classical solutions contributes to the semiclassical behavior of the vertex.

\section*{VI. Conclusions and perspectives}
In this paper we have introduced a holomorphic representation for spinfoams. The 4d spinfoam vertex for gravity has a rather elegant expression in this representation: it is given by formula (\ref{eq:WEPRL H},\ref{eq:P H}) in the Euclidean case and by (\ref{eq:Lor H},\ref{faceterm}) in the Lorentzian case. This new representation is obtained introducing first a holonomy representation and then computing its Segal-Bargmann transform. A spinfoam amplitude in the holomorphic representation is a function $W(H_l)$ of $SL(2,\mbb{C})$ elements $H_l$, one per link of the boundary graph $\Gamma$. The set of variables $H_l$ describe a truncation of the phase space of General Relativity as captured by the graph $\Gamma$. They admit a parametrization in terms of the variables generally used for describing the classical boundary geometry of a spinfoam configuration: (i) 3-normals to faces of a cellular decomposition of space, (ii) the area of faces, (iii) an angle associated to faces that measures the extrinsic curvature at the interface of two cells. These data describe the intrinsic and extrinsic geometry of the boundary and can be coded into $SL(2,\mbb{C})$ elements $H_l$ via equation (\ref{eq:Hl}). Therefore, we have a description of spinfoams in terms of variables with a clear geometrical meaning.

The holomorphic representation of spinfoams can be fruitfully used for analyzing their semiclassical behavior. In section V, we focused on a spinfoam vertex in the holomorphic representation and showed that a single exponential of the Regge action contributes to the large scale asymptotics. What selects one of the two exponentials appearing in the ``spin and normals'' representation is the peakedness of the boundary state on a prescribed extrinsic curvature. It is important to understand if this feature extends beyond the single vertex level. A way to attack the problem is to derive an action in holomorphic variables following the method introduced in \cite{Conrady:2008mk} and then checking if a mechanism of cancellation of phases as proposed in \cite{Bianchi:2008ae} is at work.

\vspace{1em}
Finally, we suggest a way to use the holomorphic representation to make a bridge with the Hamiltonian framework. Roughly, let us consider a holomorphic transition amplitude (propagation kernel) $W(H_i,H_f)$ between an initial state labeled by complex variables $H_{i}$ and a final state labeled by $H_{f}$. Suppose there is a differential operator $\hat C$ acting on either the initial or the final variables which annihilates the amplitude:
\begin{equation}
\hat C\,W(H_i,H_f)=0.
\end{equation}
The previous equation can then be interpreted as the Wheeler-DeWitt equation for pure gravity satisfied by the propagation kernel in the holomorphic representation. A recent result \cite{Bianchi:2010zs} in the context of spinfoam cosmology shows that the holomorphic amplitude for a transition between two homogeneous metrics is, in a suitable approximation, in the kernel of a differential operator which is (the quantization of) the Friedmann-Robertson-Walker Hamiltonian in the absence of matter. We find this result encouraging and it urges us to further explore this line of research.

\section*{Acknowledgments}
We thank Frank Hellmann, Roberto Pereira and Carlo Rovelli for helpful discussions. The work of E.B. is supported by a Marie Curie Intra-European Fellowship within the 7th European Community Framework Programme. E.M. gratefully acknowledges support from Fondazione A. Della Riccia. E.M. and C.P. warmly thank John W. Barrett and the European Science Foundation (ESF) for support within an Exchange Grant project.

\providecommand{\href}[2]{#2}\begingroup\raggedright\endgroup

\end{document}